\documentclass[aps,prl,twocolumn,superscriptaddress,showpacs,amsmath,amssymb,longbibliography]{revtex4-1}
\usepackage[english]{babel}
\usepackage{amsmath}
\usepackage{bm}
\usepackage{graphicx,bbm} 
\usepackage{times}
\usepackage{epsfig} 
\usepackage{soul}
\usepackage[utf8]{inputenc}
\usepackage[colorlinks,linkcolor=blue,citecolor=blue,urlcolor=blue]{hyperref}
\usepackage{color}
\usepackage{latexsym}
\usepackage{ulem}
\definecolor{nred}{RGB}{224,0,0}
\definecolor{nblue}{RGB}{28,130,185}
\definecolor{dgreen}{RGB}{78,138,21}
\definecolor{norange}{RGB}{230,120,20}
\begin{document} 
\title{Relaxation at different length-scales in models of many-body localization}
\author{J. Herbrych}
\affiliation{Department of Theoretical Physics, Faculty of Fundamental Problems of Technology, Wroc\l{a}w University of Science and Technology, 50-370 Wroc\l{a}w, Poland}
\author{M. Mierzejewski}
\affiliation{Department of Theoretical Physics, Faculty of Fundamental Problems of Technology, Wroc\l{a}w University of Science and Technology, 50-370 Wroc\l{a}w, Poland}
\author{P. Prelov\v{s}ek}
\affiliation{Jo\v{z}ef Stefan Institute, SI-1000 Ljubljana, Slovenia}
\affiliation{Faculty of Mathematics and Physics, University of Ljubljana, SI-1000 Ljubljana, Slovenia}

\date{\today}
\begin{abstract}
We study dynamical correlation functions in the random-field Heisenberg chain, which probes the relaxation times at different length scales. Firstly, we show that the relaxation time associated with the dynamical imbalance (examining the relaxation at the smallest length scale) decreases with disorder much faster than the one determined by the dc conductivity (probing the global response of the system). We argue that the observed dependence of relaxation on the length scale originates from local nonresonant regions. The latter have particularly long relaxation times or remain frozen, allowing for nonzero dc transport via higher-order processes. Based on the numerical evidence, we introduce a toy model that suggests that the nonresonant regions asymptotic dynamics are essential for the proper understanding of the disordered chains with many-body interactions.
\end{abstract}
\maketitle

\noindent {\it Introduction.} 
The phenomenon of many-body localization (MBL) deals with a challenging interplay of disorder \cite{anderson58} and interaction in many-body (MB) quantum systems \cite{basko06}, opening also fundametal questions on the statistical description of such systems. It is suggested by numerous numerical studies that prototype one-dimensional (1D) models on increasing disorder reveal the transition/crossover from an ergodic behavior to a localized regime characterized by several criteria: change in level statistics and spectral properties \cite{oganesyan07,luitz15,serbyn16,suntajs20,sierant20}, slow growth of entanglement entropy \cite{znidaric08,bardarson12,serbyn15}, vanishing dc conductivities and transport \cite{berkelbach10,barisic10,agarwal15,barlev15,steinigeweg16,prelovsek17}, nonergodic behavior of local correlations and the absence of thermalization \cite{pal10,serbyn13,huse14,luitz16,mierzejewski16}, the latter being also the experimental probe in cold-atom systems \cite{schreiber15,kondov15,luschen17}. Recently, due to the restricted system sizes available in the numerical investigations, the stability of the MBL phase has been challenged \cite{bera17,panda20}. Nevertheless, even in reachable systems, the transport as well as the relaxation properties are well defined at high temperatures $T=1/\beta \to \infty$, provided that (i) we consider properties at fixed disorder configuration, and (ii) we take into account that the frequency resolution is limited, i.e., $\delta \omega \gtrsim \omega_H \sim 1/\tau_H$, where $\tau_H$ is the Heisenberg time which in considered finite MB systems can be very long $\tau_H \propto 2^L$. 

In this work, we study the high-$T$ transport via the dynamical spin conductivity $\sigma(\omega)$, as well as local correlations embodied by the dynamical imbalance $I(\omega)$, and reveal the characteristic relaxation rates at different length-scales in the prototype model of MBL, i.e. 1D random-field Heisenberg model. While it has been already observed that the average dc value $\sigma_0=\sigma(\omega \to 0)$ depends exponentially on disorder $W$ \cite{barisic10,barisic16,steinigeweg16,prelovsek17,prelovsek21}, we establish that this is also the property for each disorder configuration. Still, differences of exponent lead to very broad (log-normal type) distribution of $\sigma_0$ \cite{barisic16,prelovsek21a} even at modest disorders $W<W_c^*$, where $W_c^*$ is the value of the presumed MBL crossover/transition $W_c^* \sim 4 J$ in the random HM \cite{pal10,luitz15,barlev15}. On the other hand, the relaxation of local quantities, as manifested in $I(\omega)$ and spatially resolved spin correlations, can reveal very small relaxation rates, which can be below the resolution $\Gamma<\delta\omega$ in considered systems, and are indication for much slower thermalization and approach to ergodicity \cite{suntajs20,Mierzejewski2021}. The observed phenomena can be well captured within a toy model, which separates for each disorder configuration the system into resonant islands \cite{thiery18,prelovsek18,sels21} and nonresonant quasi-localized islands. The transport through the latter can happen via higher-order tunneling while local thermalization occurs on much longer time scales \cite{sels21,sierant2021}. In spite of its simplicity the model accounts well for observed steep decrease of conductivity with disorder and its wide statistical spread. Similar transport properties we find also in the random transverse Ising model (TFIM) \cite{suppl}. 

\noindent {\it Model.}
In the following we mostly study the random-field Heisenberg model,
\begin{equation}
H = \sum_i \left[ \frac{J}{2}( S^+_{i+1} S^-_i + \mathrm{H.c.}) + J \Delta S^z_{i+1} S^z_i 
+ h_i S_i^z \right]\,, \label{rhm}
\end{equation}
with spin $S=1/2$ operators and $\Delta=1$, while $h_i\in[-W,W]$ are local fields with uniform probability distribution. We consider 1D chains with $L$ sites and periodic boundary conditions, with $J=1$ as the energy unit. We first concentrate on the high-$T$ ($T \gg J$) dynamical spin conductivity,
\begin{equation}
\tilde\sigma(\omega) = T \sigma(\omega) = \frac{1}{L} \int_0^\infty \mathrm{d}t\, \mathrm{e}^{i \omega t} 
\langle j(t) j \rangle\,, \label{sig}
\end{equation}
related to the uniform spin current \mbox{$j = (J/2) \sum_j( i S^+_{j+1} S^-_j + \mathrm{H.c.})$}. We calculate $\tilde\sigma(\omega)$ (and other dynamical correlation functions considered in this work) for each disorder configuration using upgraded microcanonical Lanczos method (MCLM) \cite{long03,karahalios09,prelovsek11,prelovsek21} with high-resolution $\delta \omega$. The method evaluates the dynamical correlations within a microcanonical state $|\Psi_{\cal E}\rangle$ corresponding to chosen energy $\cal E$ (which we choose here in the middle of the MB spectrum, i.e., ${\cal E} \sim 0$) and with small energy dispersion $\sigma_{\cal E} < \delta \omega \sim \Delta E /M_L$, obtained via large number of Lanczos iterations $M_L$, where $\Delta E$ is the system MB energy span. In the following we present results for $L=26$ sites in the $S^z_{tot}=0$ sector, with the number of MB states $N_{st} \sim 10^7$ states, and by using $M_L \sim 2.10^5$ we reach (in considered disorder range $W \leq 4$) the resolution $\delta \omega \sim 10^{-4}$, still larger than $\omega_H\sim\Delta E/N_{st}\lesssim 10^{-5}$. 

\begin{figure}[tb]
\includegraphics[width=1.0\columnwidth]{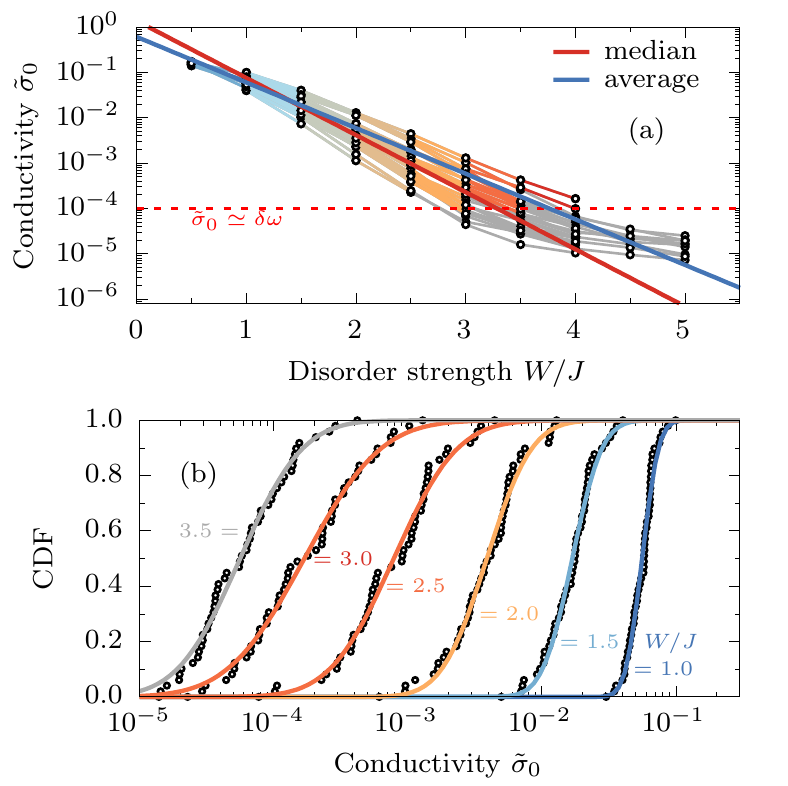}
\caption{(a) High-$T$ dc spin conductivities $\tilde\sigma_0$ within the random-field Heisenberg model vs. disorder strength $W$ for different disorder realizations (30 samples for each $W$), as evaluated with MCLM on $L=26$ system. The thick lines represent exponential fits to the median and average $\tilde\sigma_0$. (b) Cumulative distribution function (CDF) of $\tilde\sigma_0$ values for different $W$ (50 samples for each $W$). Curves represent log-normal distributions as the guide to the eye.}
\label{fig1}
\end{figure}

It should be noted that even within a finite system for chosen sample $h_i$ and energy $\cal E$ dynamical $\tilde\sigma(\omega)$, and in particular $\tilde\sigma_0$, are well defined and resolved provided that $\tilde\sigma_0\gtrsim\delta\omega$ (see \cite{suppl} for typical spectra $\tilde\sigma(\omega)$ at different $W$). In Fig.~\ref{fig1}(a) we summarize results for dc $\tilde\sigma_0$ at increasing $W$, where we choose $h_i=W\eta_i$ with random configurations $\eta_i\in[-1,1]$. We note that $W\sim 1$ roughly represents \cite{suppl} the borderline between the weak scattering regime and the incoherent diffusion where $\sigma(\omega)$ is maximum at $\omega>0$. Results in Fig.~\ref{fig1}(a) generally reveal for $W>1$ an exponential-like dependence $\tilde\sigma_0\propto\exp(-b W)$ for each disorder configuration separately, with typical $b\sim 2.5$ \cite{barisic10,steinigeweg16,barisic16,prelovsek17}. Still, slightly different (sample dependent) $b$ lead to a large statistical spread of $\tilde\sigma_0$ value, as summarized in Fig.~\ref{fig1}(b) by the cumulative distribution function (CDF) which is close to the log-normal distribution. We also note that our $\delta\omega$ resolution limits reliable $\tilde\sigma_0\gtrsim10^{-4}$ for different samples at marginal $W^*_c \sim 4 $, with the variation $\delta W^*_c\sim 0.5$.

\begin{figure}[tb]
\includegraphics[width=1.0\columnwidth]{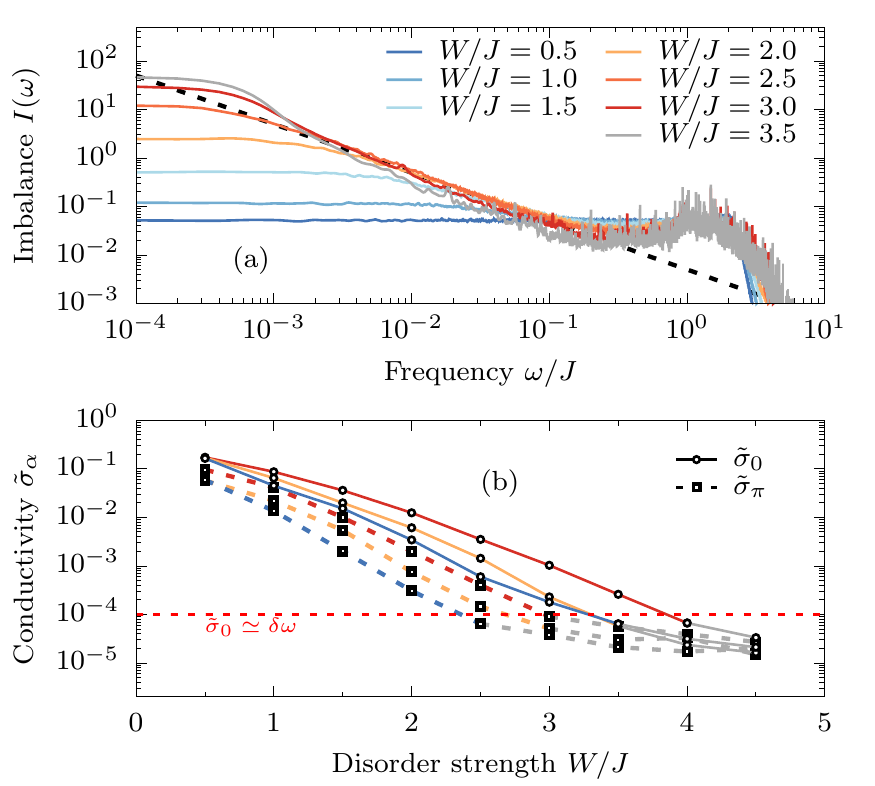}
\caption{(a) Dynamical imbalance $I(\omega)$ within a single disorder sample at different strength $W$, calculated for $L=26$. Dashed line depict $1/\omega$ dependence. (b) Extracted dc spin conductivity $\tilde\sigma_\pi$ in comparison with the uniform $\tilde\sigma_0$ vs. $W$ for three potential configurations.}
\label{fig2}
\end{figure}

In contrast to $\tilde\sigma_0$, which can be typically well followed for $W \lesssim 3.5$, local correlations can reveal already exceedingly long relaxation times. Of interest is the spin correlations $S_q(\omega)$ of modulation operator $S^z_{q} = (1/\sqrt{L}) \sum_j \exp(iq j) S^z_j$ which can be related \cite{herbrych2012,mierzejewski16,prelovsek17-2,prelovsek17,suppl} to the $q$-dependent spin conductivity $\tilde\sigma_q$ 
\begin{equation}
S_q(\omega)=-\frac{1}{\pi} \mathrm{Im} \left[ \frac{\chi_q^0}{\omega+i\ g_q^2\tilde\sigma_q(\omega)/\chi_q^0} \right]\,.
\end{equation}
Here $g_q=2\sin(q/2)$, and $\chi_q^0=\langle S^z_{-q} S^z_q \rangle=1/4$. Note also that $S_\pi(\omega)=I(\omega)$ is directly relevant to cold-atom experiments \cite{schreiber15,kondov15,luschen17}, i.e., $I(\omega)$ probes the local thermalization, in particular the relaxation rate $\Gamma_I \propto \tilde\sigma_{\pi}(\omega \to 0)=\tilde\sigma_{\pi}$, determined by the saturation $I(\omega < \Gamma_I) \propto \tilde 1/\sigma_{\pi} $. Results in Fig.~\ref{fig2}(a) for a single disorder configuration, reveal that $\Gamma_I$ can become very small and hardly resolved in considered system, i.e., $\Gamma_I \lesssim \delta\omega$, even at modest $W \sim 2.5$, where $\tilde\sigma_0$ is still well defined. An indication of finite-size dominated $\Gamma_I$ is also the deviation from marginal $I(\omega) \propto 1/\omega$ \cite{mierzejewski16,prelovsek17} at larger $W \lesssim W^*_c$. Furthermore, the results for $\tilde\sigma_\pi$ extracted from $I(\omega)$ for few samples are presented in Fig.~\ref{fig2}(b) and confirm than in general $\tilde\sigma_\pi<\tilde\sigma_0$, with the difference becoming large on approaching $W\sim W_c^*$, i.e., indicated increasing difference between thermalization and (local) transport relaxation.

\begin{figure}[tb]
\includegraphics[width=1.0\columnwidth]{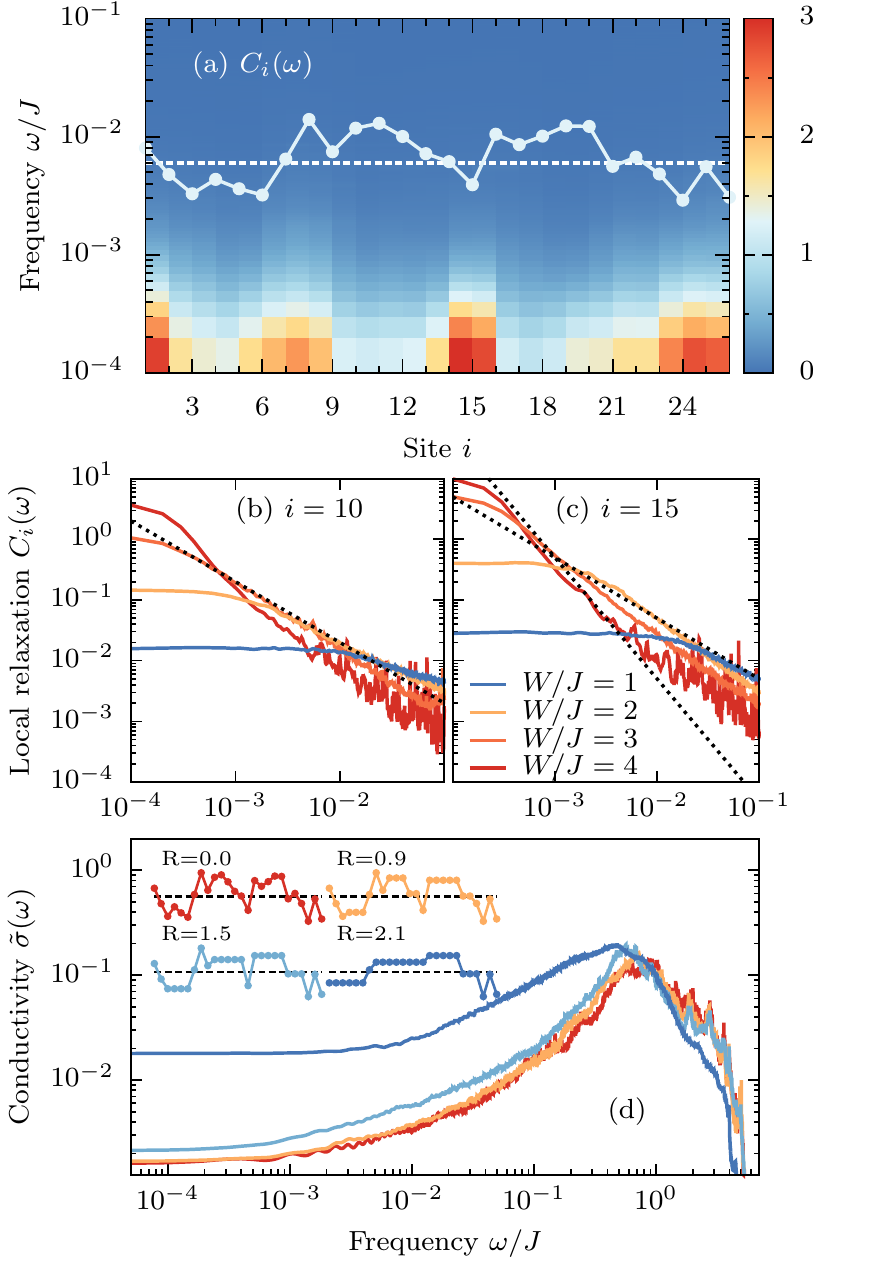}
\caption{(a) Local spin correlations $C_i(\omega)$ for all sites $i=1,L$ with potentials $h_i$ corresponding to $W=3$ (with the potential landscape also shown). (b,c) Local correlations $C_i(\omega)$ for different $W=1-3$ for two characteristic sites $i=10,15$, representing resonant and localized islands, respectively. Dashed line in (b) represent $\propto1/\omega$ dependence, while in (c) $\propto1/\omega$ and $\propto1/\omega^2$. (d) $\tilde\sigma(\omega)$ for flattened potentials with various thresholds $R=0-2.1 $ for one configuration with $W=3$.}
\label{fig3}
\end{figure}

While $I(\omega)$ and related $\Gamma_I$ monitor the local relaxation averaged over all sites in the system, it is instructive to follow also the local $C_{i}(\omega)$, i.e., correlations of $S^z_i$ for each site in the chain. In Fig.~\ref{fig3}(a) we present a variation of $C_i(\omega)$ among all sites in one chosen configuration at moderate $W=3$. We note substantial variations in low-frequency $C_i(\omega \to 0) \propto 1/\Gamma_i$, whereby small $\Gamma_i$ can be directly correlated with large potential deviations of local $h_i$, also presented in Fig.~\ref{fig3}(a). More detailed comparison of $C_i(\omega)$ for two typical sites, representing the weak potential fluctuation, site $i=10$, and strong potential-fluctuation regime at $i=15$, respectively, are shown in Figs.~\ref{fig3}(b,c). While generally $C_{15}(\omega\sim 0)\gg C_{10}(\omega \sim 0)$, it also appears that $C_{15}(\omega)$ reveals for $W\geq 3$ [instead of $C_i(\omega>\Gamma_i)\propto 1/\omega$] a Lorentzian behavior \cite{Mierzejewski2021} with $C_i(\omega>\Gamma_i) \propto 1/\omega^2$, representing the marginal $\Gamma_i \sim \delta \omega$. It follows from Fig.~\ref{fig3}(a) that the local dynamics captured in $C_i(\omega )$ is particularly slow in the vicinity of sites $i$ with large difference in local potentials $|h_i-h_{i\pm 1}|$. One could in fact expect for the latter a Lorentzian with $\Gamma_i\propto\exp(-a |h_i|)$, with large $a>3 $ \cite{sels21}. While we simulate such case in \cite{suppl}, we find much smaller $a \sim 1$, which can better account for observed differences in $\Gamma_i$. 

As the main result we demonstrate that the regions with slow dynamics are essential not only for thermalization of local operators but also for transport at various length-scales, i.e., $\tilde{\sigma}_0$ vs. $\tilde{\sigma}_\pi$, and various time-scales. We introduce a parameter $R$ which allows to distinguish between the {\it nonresonant (localized) } islands and the {\it resonant (conducting)} islands. Namely, we assume that site $i$ belongs to an isolated island when $|h_i-h_{j}| > R$ for both neighboring sites, $j=i\pm1$. Otherwise, $i$ belongs to a conducting island. We note that taking the standard resonant scenario \cite{prelovsek18,prelovsek21}, together with the matrix elements relevant for the spin-flip $J/2$, one gets $R = J = 1$. Interestingly, we observe that the spatial-variation of $h_i$ within the conducting islands is not essential for transport. To this end, for all sites $i$ which belong to the conducting islands we replace $h_i$ with $\bar{h}_i$, where $\bar{h}_i$ is the average $h_j$ over all sites $j$ in the same island. Such flattening eliminates disorder within each conducting island, as it is sketched in Fig.~\ref{fig3}(d) for a single configuration at $W=3$. We notice from Fig.~\ref{fig3}(d) that the resulting $\tilde\sigma(\omega)$ is hardly affected up to $R\simeq1.5$. This result reveals a clear separation of the studied system into conducting and localized islands, and shows that the transport is determined by the localized islands. 

\noindent {\it Toy model.}
Since the time scales which are relevant for the dynamics in the conducting and isolated islands differ substantially, the latter can be considered to be frozen and transformed out from the Hamiltonian. Then, the transport through a localized island that contains $M$ frozen spins, $S^z_{j}\cdots S^z_{j+M-1}$, can happen via high-order virtual process involving at least $M$ spin flips. It leads to new effective spin-flip term, $\tilde H'_j=(J_j^{\mathrm{eff}}/2)(S^+_{j+M} S^-_{j-1} + \mathrm{H.c})$, between sites $j-1$ and $j+M$, which belong to the neighboring conducting islands \cite{sels21}. One can derive $J_j^{\mathrm{eff}}$ via the $M$-th order degenerate perturbation theory
\begin{equation}
\tilde H'_j = H' Q \frac{1}{\bar E -H_0} H' Q \cdots Q H'\frac{1}{\bar E -H_0} Q H'\,, \label{dpt}
\end{equation}
where $H_0=H_h+H_\Delta$ and $Q$ projects all intermediate states equal to the initial or the final ones. Here, $H'$, $H_\Delta$ and $H_h$ denote, respectively, the first, second and the last term in the Hamiltonian~\eqref{rhm}. In order to obtain analytical estimate for $J_j^{\mathrm{eff}}$, we introduce few simplifications. We assume strong disorder ($H_0 \simeq H_h$) ferromagnetic states of isolated islands ($S^z_{i}=S^z_{i'}$ for $i,i'=j,...,j+M-1$) and we fix $\bar E= (E_i+E_f)/2 $ as the average between initial and final $H_h$, taking also $\bar h_{j-1} \sim \bar h_{j+M} \sim 0$. Then, one can directly evaluate Eq.~\eqref{dpt} and the effective coupling
\begin{equation}
|J^{\mathrm{eff}}_j| = \frac{J^{M+1}}{2^M}\frac{1}{h_j h_{j+1}\cdots h_{j+M-1}}\,.
\label{jeff}
\end{equation}
It is interesting to note that Eq.~\eqref{dpt} remains valid for other spin configurations apart from ferromagnetic, what we have checked explicitly for $M\le 3$.

It is now straightforward to define a toy model which relies on the assumption that dc transport is dominated by the {\it incoherent} conduction via localized islands. Then, the transport appears through the sequence of incoherent hoppings (series of resistors), i.e., corresponding to $J_i=J$ for link in the resonant islands and $J_i=J^{\mathrm{eff}}_j/M$ for each link in the nonresonant ones. Under such conditions, the conductivity is given by $\tilde\sigma_0\sim\sigma^*_0 \bar J^{\mathrm{eff}}$ and $1/\bar J^{\mathrm{eff}} = (1/L)\sum_i(J_i)^{-1}$ where the numerical constant $\sigma^*_0 \sim 0.1$ is chosen to reproduce the incoherent dc conductivity at $W\sim1$. The perturbative expression, Eq~\eqref{jeff}, is applicable for realizations of disorder where potentials inside the island are not degenerate with the outside ones, i.e., $|h_i|\gg 0$. In order to account for the latter, we neglect localized islands with $M$ sites for which $J^{\mathrm{eff}}_{i}>J/M$.

\begin{figure}[tb]
\includegraphics[width=1.0\columnwidth]{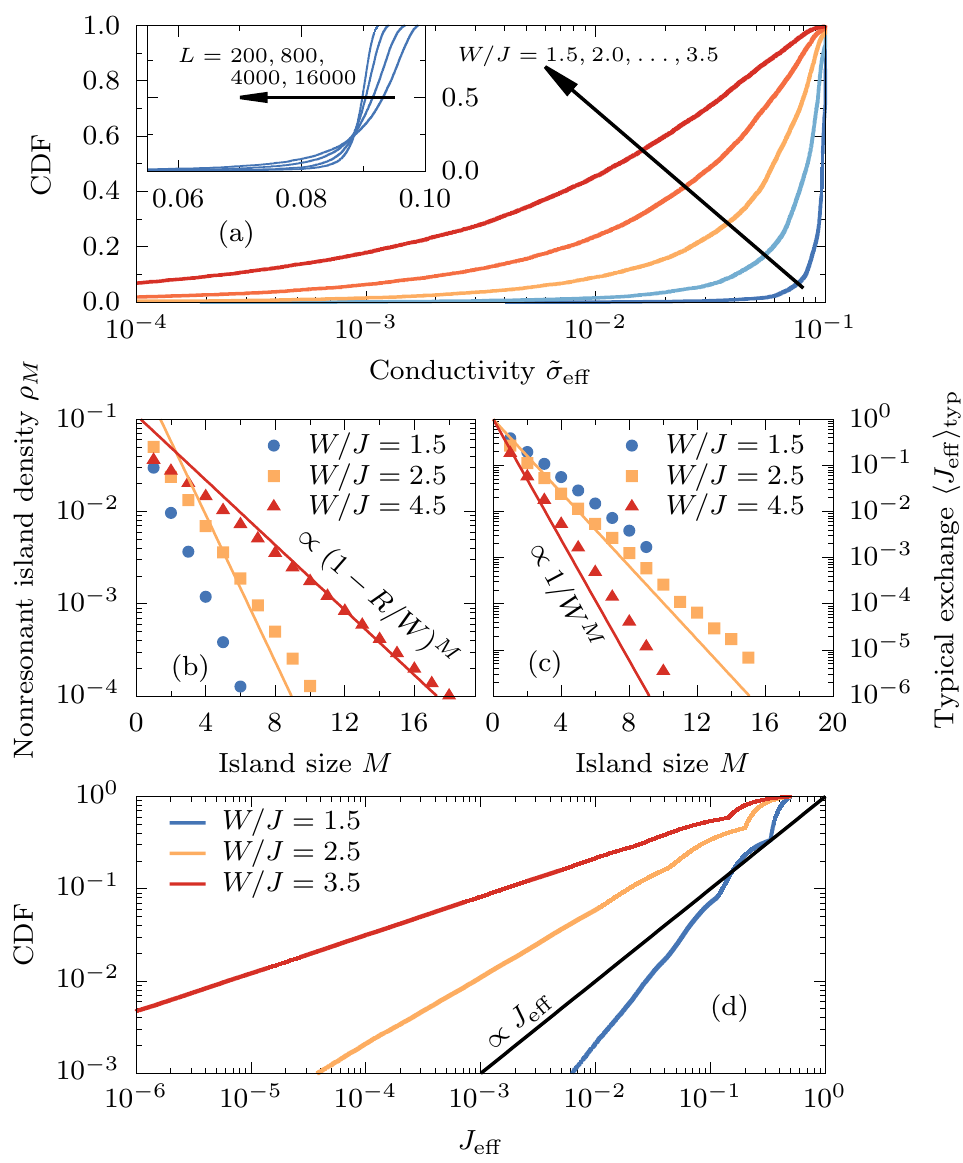}
\caption{(a) Toy-model dc conductivity $\tilde\sigma_0$ for various $W$ and $L=25$. The inset show $L=20-16000$ dependence for $W=1.5$. (b),(c) Density $\rho_M$ and typical value of $J^{\mathrm{eff}}$ for nonresonant islands of length $M$ obtained for $L=10^7$. (d) Distribution of $J^{\mathrm{eff}}$ in the toy-model}.
\label{fig4}
\end{figure}

We have carried out simulations of the toy model following all steps previously tested via full quantum calculations. For each realization of disorder we identify the resonant and nonresonant islands, flatten the disorder within the resonant islands ($h_i\to\bar{h}_i$) and evaluate $\tilde\sigma_0$. Based on results shown in Fig.~\ref{fig3}(d) we take $R=1.5$. The main panel in Fig.~\ref{fig4}(a) shows the CDF of $\tilde\sigma_0$ obtained for different disorder realizations. The toy model correctly reproduces the main features of the full-quantum calculations. In particular, the median of the CDF decays approximately exponentially with $W$, however the rate is slower than results shown in Fig.~\ref{fig1}(b). Moreover, the distribution of $\tilde\sigma_0$ for $L\sim25$ is very broad indicating that the spread of $\tilde\sigma_0$ may span over a few orders of magnitude. For weak disorder, the width of the distribution decreases with $L$ and for sufficiently large systems the distribution approaches the normal (Gaussian) distribution expected for diffusive systems, as shown in the inset in Fig.~\ref{fig4}(a).

For stronger disorder, the toy model reveals an anomalous Griffiths-scenario reported in several numerical studies \cite{agarwal15,gopal16,mierzejewski20}. To explain its origin, in Fig.~\ref{fig4}(b) we show the density of nonresonant islands of length $M$, $\rho_M=N_M/L$, where $N_M$ denotes the number of localized islands and the simulations were carried out for $L=10^7$.  The probability of finding a nonresonant link in strongly disordered system is $(1-R/W)$, hence $\rho_M\propto (1-R/W)^M$. Fig.~\ref{fig4}(c) shows the typical value of $J_{\mathrm{eff}}$ for nonresonant islands of length $M$. We find $\langle J_{\mathrm{eff}} \rangle_{\rm typ}\propto W^{-M}$, what straightforwardly follows from Eq.~\eqref{jeff}. The exponential dependence of $\rho_M$ and $J_{\mathrm{eff}}(M)$ on $M$ is very robust. As consequence, the toy model realizes the Griffiths scenario, i.e., large nonersonant islands are exponentially rare but the corresponding $1/J_{\mathrm{eff}}$ is exponentially large. Consequently, such islands have substantial impact on transport. The interplay leads to a power-law $\mathrm{CDF}\propto J_{\mathrm{eff}}^\alpha$ shown in Fig.~\ref{fig4}(d). The probability density $f(J_{\mathrm{eff}}) \propto J^{(\alpha-1)}_{\mathrm{eff}}$ and the average $\langle J^{-1}_{\mathrm{eff}} \rangle=\int_{0}^{1} \mathrm{d} J\; f(J)/J $ is finite only for $\alpha > 1$ and diverges otherwise. The latter implies $\tilde\sigma_0 \to 0$ in the $L \to \infty$ limit, indicating subdiffusive transport or localization. It should be stressed that the toy model is based on the assumption that the nonresonant islands are strictly frozen. If the lifetimes are finite, then the nondiffusive transport may be transient while the asymptotic transport may be still diffusive. It is also possible that the role of large islands may be overrepresented due to possible inherent internal resonances. Results in Fig.~\ref{fig3}(c) indicate that the relaxation of the nonresonant islands can be studied numerically only up to relatively weak disorder $W \simeq 2.5$.

\noindent {\it Conclusions. } We have studied dynamical correlation functions which probe the relaxation mechanisms at different length-scales in the random-field Heisenberg model. The dc conductivity $\tilde\sigma_0$ probes the transport at large length-scales (wave-vectors $q \to 0$) and was shown to decrease exponentially with disorder with an increasing log-normal-like distribution of $\tilde \sigma_0$ values. Still, its decrease disorder is much slower than the dynamical imbalance and the corresponding $\tilde\sigma_{\pi}$, which probes the smallest length-scales ($q =\pi$). We argue that the surprising difference between $\tilde\sigma_{0}$ and $\tilde\sigma_{\pi}$ originates from the presence of nonresonant regions (islands) with particularly slow or completely frozen dynamics. These extremely long (or infinite) relaxation times are probed by the long-time dynamics of the imbalance or, equivalently, by $\tilde\sigma_{\pi}(\omega \to 0)$. However, their contribution to $\tilde\sigma_0$ can be treated perturbatively in that spins excitation can pass nonresonant islands via virtual spin-flip processes (on the much shorter time scales). As a consequence the nonresonant islands influence $\tilde\sigma_{\pi}$ more strongly than $\tilde\sigma_{0}$. Still, the presence of nonresonant islands explains large sample-to-sample spread of $\tilde\sigma_0$ and the exponential dependance of $\tilde\sigma_0$ on the strength of disorder. Large nonresonant islands may give rise also to anomalous (nondiffusive) transport for stronger disorder, although their role can be overestimated in the present study. It is evident from the presented results that the fate of the MBL phase and transport properties of the disorder many-body systems depends on the relaxation times of the nonresonant islands, i.e., whether the latter are finite or infinite.

\begin{acknowledgments}
J.H. acknowledges the support by the Polish National Agency of Academic Exchange (NAWA) under contract PPN/PPO/2018/1/00035. M.M. acknowledges the support by the National Science Centre, Poland via projects 2020/37/B/ST3/00020. P.P. acknowledges the support by the project N1-0088 of the Slovenian Research Agency. The numerical calculation were partly carried out at the facilities of the Wroclaw Centre for Networking and Supercomputing.
\end{acknowledgments}

\bibliography{manurelax}
\newpage
\phantom{0}
\newpage
\setcounter{figure}{0}
\setcounter{equation}{0}
\setcounter{page}{0}

\renewcommand{\thetable}{S\arabic{table}}
\renewcommand{\thefigure}{S\arabic{figure}}
\renewcommand{\theequation}{S\arabic{equation}}
\renewcommand{\thepage}{S\arabic{page}}

\renewcommand{\thesection}{S\arabic{section}}

\onecolumngrid

\begin{center}
{\large \bf Supplemental Material:\\
Relaxation at different length-scales in models of many-body localization}\\
\vspace{0.3cm}
J. Herbrych$^{1}$, M. Mierzejewski$^{1}$ and P. Prelov\v{s}ek$^{2,3}$\\
$^1${\it Department of Theoretical Physics, Faculty of Fundamental Problems of Technology, \\ Wroc\l aw University of Science and Technology, 50-370 Wroc\l aw, Poland}\\
$^2${\it Department of Theoretical Physics, J. Stefan Institute, SI-1000 Ljubljana, Slovenia} \\
$^3${\it Department of Physics, Faculty of Mathematics and Physics, University of Ljubljana, SI-1000 Ljubljana, Slovenia}
\end{center}

\vspace{0.6cm}

\twocolumngrid

\label{pagesupp}

\section{General structure of dynamical conductivity $\tilde\sigma(\omega)$} \label{app1}

\begin{figure}[!htb]
\includegraphics[width=1.0\columnwidth]{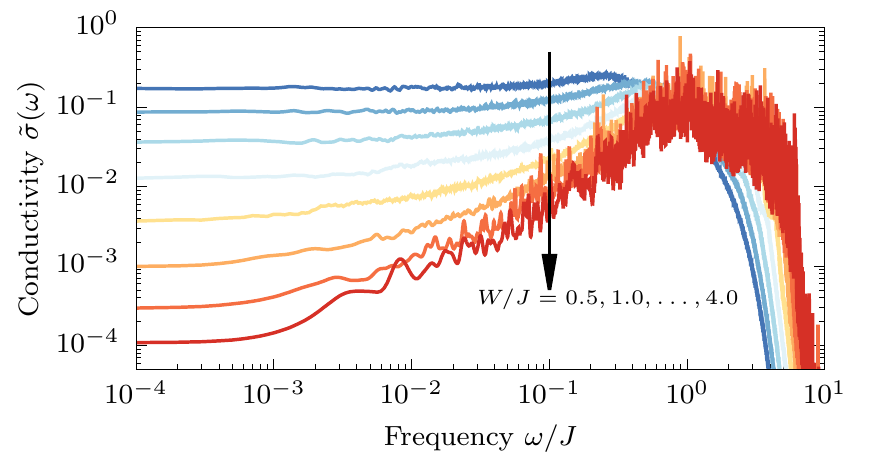}
\caption{High-$T$ dynamical spin condictivity $\tilde\sigma(\omega)$ as calculated with MCLM at fixed energy ${\cal E} = 0$ for random-field Heisenberg model on $L=26$ sites for selected potential distribution but different disorders $W/J = 0.5 - 4.0$.}
\label{figS1}
\end{figure}

It is instructive to follow the whole spectrum of high-$T$ dynamical conductivity $\tilde\sigma(\omega)$ within the random-field Heisenberg model (HM), as this evolves with the increasing disorder $W$. In Fig.~\ref{figS1} we present the typical result for a single random potential configuration (and target energy ${\cal E} = 0$), as obtained using MCLM on a system of $L=26$ sites. While it has been established that close to the crossover $W\sim W^*_c$ the conductivity follow $\tilde\sigma(\omega\to0)\sim\tilde\sigma_0+c|\omega|^\alpha$ with $\alpha \sim 1$ \cite{karahalios09,barisic16,steinigeweg16}, the log-plot presentation in Fig.~\ref{figS1} enhances the low-$\omega$ part and the importance of high frequency resolution $\delta\omega\sim 10^{-4}$ in the present study, allowing to follow dc conductivity numerically down to $\tilde \sigma_0 \sim \delta \omega$, and consequently up to $W \sim W_c^* \lesssim 4$. Note that, since we study isotropic HM there is no coherent (dissipationless) transport even at $W=0$, still finite disorder starts to dominate $\tilde\sigma(\omega)$ in the incoherent regime $W> W^* \sim 1$ where also spectral maximum is clearly at $\omega_M>0$.

\section{Dynamical imbalance and $q$-dependent conductivity} \label{app2}

General dynamical spin correlations are defined by 
\begin{equation}
S_q(\omega)=\frac{1}{\pi} \int_0^\infty \mathrm{d}t \mathrm{e}^{i \omega t} \langle S^z_{-q}(t) S^z_{q} \rangle\,,
\end{equation}
where $S^z_{q}=(1/\sqrt{L})\sum_i\mathrm{e}^{i q j} S^z_j$. For $q=\pi$ the quantity corresponds to imbalance correlations $I(\omega)$. Furthermore, one can define (for on average homogeneous system) the corresponding $q$-dependent spin conductivity $\tilde\sigma_q(\omega)$ via the general representation of the corresponding $S^z_{q}$ (complex) relaxation function $\phi_q(\omega)$ \cite{herbrych2012,mierzejewski16,prelovsek17-2}, where we again deal with $\beta =1/T\to 0$ case,
\begin{equation}
\phi_q(\omega)=\frac{-\beta\chi_q^0}{\omega+M_q(\omega)},\qquad M_q(\omega)=
i\frac{g_q^2}{\chi_q^0}\tilde\sigma_q(\omega)\,.
\end{equation}
Here $\mathrm{Im} \phi_q(\omega) = \pi \beta S_q(\omega)$, $g_q=2\sin(q/2)$, and $\chi_q^0=\langle S^z_{-q} S^z_q \rangle=1/4$. In general one expects $\tilde\sigma_{q\to0}(\omega)=\tilde\sigma(\omega)$ although numerically this is hard to follow in accesible finite-size systems. 

\section{Local dynamical spin correlation at localized site} \label{app3}

It is instructive to understand the simplified problem of local spin correlation $C_i(\omega)$ in the case, when a single site $i_0$ has larger potential $V \gg J$, while the rest of the system has only a weak disorder $W \leq J$. Here, the background disorder has the role to avoid the pathologies of the clean (integrable) model. Such a scenario has been recently considered \cite{sels21} and the (time-dependent) spin correlations on site $i = i_0$ were predicted to show, at $V>1$, an exponential decay $C_{i_0}(t)\propto\exp(-\Gamma t)$ with the characteristic rate $\Gamma\propto\exp[-\alpha V \log(V)]$, exponentially dependent on $V$ due to multi-magnon processes necessary to absorb the local spin-flip energy $\propto V$.

\begin{figure}[tb]
\includegraphics[width=1.0\columnwidth]{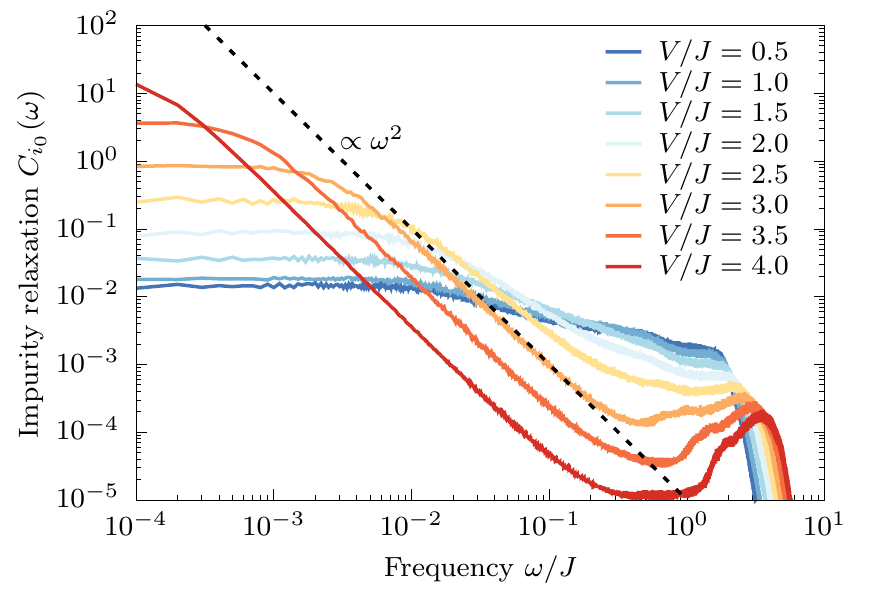}
\caption{Local dynamical spin correlations $C_{i_0}(\omega)$ for a spin on a site with large potential $V \geq 1$ while the background has weak disorder $W=1$. Calculated via MCLM on a system with $L=26$ sites.}
\label{figS2}
\end{figure}

We explicitly test this scenario by performing a MCLM calculation of local $C_{i_0}(\omega)$ on a system of $L=26$ sites, by varying local potential $V/J$ while keeping the background with fixed and modest disorder $W = J$. We first note that the results are rather insensitive to background disorder provided that disorder is modest $W \leq J$ (but still not too small). Again, it is crucial to have high frequency resolution $\delta \omega \ll J$ to distinguish Lorentzian $C_{i_0}(\omega)$ from a $\delta(\omega)$ contribution, which represents the finite-size limitation in this problem. Results in Fig.~\ref{figS2} indeed confirm that we calculation well captures the regime $V \leq 4$ where $C_{i_0}(\omega)$ is Loretzian with exponential-like dependence of $\Gamma\propto\exp(-a V)$. However, in contrast to prediction \cite{sels21} where $ a \sim 3\log(V)\gg 1$ we find much more modest $ a \sim1$. 

\section{Random transverse-field Ising model} \label{app5}

Another 'standard' model for MBL is 1D random transverse-field Ising model (TFIM),
\begin{equation}
H = \sum_i [ J_i \sigma_i^z \sigma^z_{i+1} + h_i \sigma^z_i + \gamma_i \sigma^x_i ] = \sum_i H_i~\,,
\end{equation}
where $\sigma$ are Pauli matrices. As in several previous studies \cite{roy192,roy21,abanin21}, we adopt uniform $\gamma_i=\gamma$, random $h_i=[-\tilde W,\tilde W] $, and $J_i=J+\delta J_i$ with $\delta J_i=[-W_J,W_J]$ and taking $J=1$ as the unit energy. Since in TFIM $S^z_{\mathrm{tot}}$ is not conserved, the only relevant transport is of energy density $H_i$ via the energy current
\begin{equation}
j_E=-\frac{1}{2} \sum_i J_i [\gamma_{i+1}\sigma^z_i\sigma^y_{i+1}-\gamma_{i}\sigma^y_i\sigma^z_{i+1}]\,.
\end{equation}
Using $j^E$ high-$T$ dynamical energy conductivity $\tilde\sigma^E(\omega)$ can be defined in analogy with Eq.~\eqref{sig}, whereby it is related (at $T \gg J$) to standard (dynamical) thermal conductivity as $\kappa(\omega)=\tilde\sigma^E(\omega)/T^2$. 

In Fig.~\ref{figS2}(a) we present typical results for $\tilde\sigma^E(\omega)$, as calculated using MCLM on a system with $L=22$ (smaller then in the case of HM which has conserved $S^z_{\mathrm{tot}}$, but still $N_{st} \sim 4.10^6$) for the standard choice of parameters, i.e., fixed $\gamma=J$, modest $W_J=0.2$, and varying local-field disorder 
$\tilde W=1-4$. Results for $\tilde\sigma^E(\omega)$ are qualitatively similar to $\sigma(\omega)$ in the random HM (compare with Fig.~\ref{figS1}). Note that the energy span of the TFIM is much larger than in the HM, $\Delta E\sim10^2$, leading also to smaller frequency resolution (taking $M_L\sim10^5$ leads typically to $\delta\omega\sim10^{-3}$). It is evident from the presented results that the incoherent transport of TFIM sets in at $W^* \sim 1.5$, where $\tilde\sigma^E(\omega)$ becomes featureless up to the scale $\omega \sim 1$, while for $W>W^*$ the maximum $\tilde\sigma^E(\omega)$ moves to $\omega_M \sim 1$. At the same time, the variation of dc values $\tilde\sigma^E_0$ is exponential-like for each disorder sample [see Fig.~\ref{figS1}(b)], and the statistical distribution is becoming very wide and log-normal type [similarly as in HM shown in Fig.~\ref{fig1}(b) of the main text]. Taking into account the MCLM resolution, $(\tilde \sigma^E_0)_{min}\sim 10^{-3}$, we can follow the steady exponential decrease down to $W\sim W_c^*\sim 4$, which is beyond - but still consistent with - previous findings of the transition/crossover $W^*_c\sim 3.5$ \cite{abanin21,roy21}.

\begin{figure}[tb]
\includegraphics[width=1.0\columnwidth]{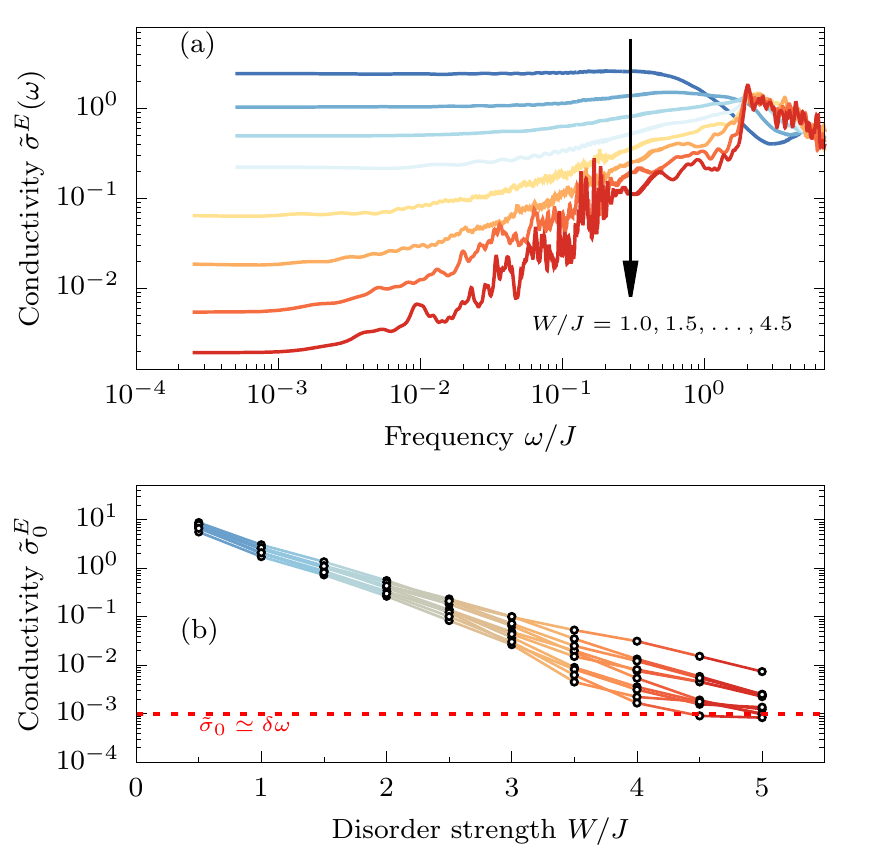}
\caption{High-$T$ dynamical energy conductivity $\tilde\sigma^E(\omega)$ as calculated with MCLM at fixed energy ${\cal E} = 0$ for random TFIM on $L=22$ sites for different disorders $W = 1.0 - 4.0$. (b) The energy conductivity $\sigma^E_0$ vs. disorder $W$ within the random TFIM for five different potential configurations.}
\label{figS3}
\end{figure}

TFIM results for $\tilde\sigma^E(\omega)$ and $\tilde\sigma^E_0$ as the function of disorder $W$, reveal close similarity with the corresponding results for $\tilde\sigma(\omega)$ with the random-field HM. One clear difference is that random TFIM (in contrast to $\Delta\to0$ case for the anisotropic HM) does not have any limit which would correspond to the Anderson model of noninteracting particles \cite{anderson58}. To find a closer relation with random-field HM, one can perform local spin rotations (in the $S=1/2$ representation) on each site $i$, $ \tilde H = R^{-1} H R$ with $R = \prod_i R_i$, so that local fields turn into $z$ direction, i.e., for rotation angle $|\phi_i| < \pi/2$, 
\begin{equation}
R_i=\cos\frac{\phi_i}{2}-2iS^y_i\sin\frac{\phi_i}{2}\,, \quad \tan\phi_i=\epsilon_i=\frac{\gamma}{|h_i|}\,.
\end{equation}
Transformed $\tilde H=\tilde H_z+\tilde H_{xx}+\tilde H_{x}$ can be expressed as,
\begin{eqnarray}
\tilde H_z&=&2 \sum_i[\tilde h_i S^z_i+ 2 J_i c_i c_{i+1} S^z_i S^z_{i+1} ]\,, \nonumber \\
\tilde H_{xx}&=&4\sum_i J_i s_i s_{i+1} S^x_i S^x_{i+1}\,, \\
\tilde H_{x}&=&-4\sum_i s_i S^x_i [J_{i-1} c_{i-1} S^z_{i-1} +J_i c_{i+1} S^z_{i+1}]\,, \nonumber
\end{eqnarray}
where $\tilde h_i=h_i\sqrt{1+\epsilon_i^2}$ and $c_i=\cos\phi_i\,,~s_i= \sin\phi_i$. The main difference to the anisotropic HM is that $\tilde H$ does not conserve total spin. Recovering for convenience spin rotation ($x-y$ symmetry) and consequently $S^z_{tot}$ conservation, neglecting $\tilde H_x$ representing random (sign)
interaction term, as well as averaging the interaction terms (where coefficients are all positive), we get
\begin{equation}
\tilde H \sim 4 \sum_i [\frac{\tilde h_i}{2} S^z_i+ J_i \bar c^2 S^z_i S^z_{i+1} + \frac{ J_i \bar s^2}{2} (S^x_{i+1} S^x_i
+ S^y_{i+1} S^y_i) ]\,.
\end{equation} 
In our study (as well as in most previous ones) $W \ll1$ while $W\gg\gamma$, so that the correspondence to the anisotropic HM is $\bar c\sim1$ and $\bar s\propto\gamma/W\ll1$. As a consequence, the effective anisotropy (relative to the effective exchange $J_{\mathrm{eff}} = \bar s^2/2$) is given by $\Delta_{\mathrm{eff}}=2 \bar c^2/\bar s^2\gg1$. Within the interesting regime of random TFIM ($\gamma=1$ and $W_J\ll1$), the main difference is in $\Delta_{\mathrm{eff}}\gg1$, which pushes up also the effective critical $w^*_c=W^*_c/J_{\mathrm{eff}}$ to much higher value relative to the corresponding (most studied isotropic) random-field HM. Still, the general properties (as, e.g., transport dc quantities) apparently behave qualitatively similar in both models.

\end{document}